# Hierarchical structure and biomineralization in cricket tooth*

XING Xue-Qing (邢雪青) [1,2]   GONG Yu (宫宇) [1,2]   CAI Quan (蔡泉) [1]
MO Guang (默广) [1]   DU Rong (杜蓉) [1,2]   CHEN Zhong-Jun (陈中军) [1]
WU Zhong-Hua (吴忠华) [1,1)]

[1]. Institute of High Energy Physics, Chinese Academy of Sciences, Beijing 100049, China
[2]. Graduate University of the Chinese Academy of Sciences, Beijing 100049, China

**Abstract:** Cricket is a truculent insect with stiff and sharp teeth as a fighting weapon. The structure and possible biomineralization of the cricket teeth are always interested. Synchrotron radiation X-ray fluorescence, X-ray diffraction and small angle X-ray scattering techniques were used to probe the element distribution, possible crystalline structures and size distribution of scatterers in cricket teeth. Scanning electron microscope was used to observe the nanoscaled structure. The results demonstrate that Zn is the main heavy element in cricket teeth. The surface of the cricket teeth has a crystalline compound like $ZnFe_2(AsO_4)_2(OH)_2(H_2O)_4$. While, the interior of the teeth has a crystalline compound like $ZnCl_2$, which is from the biomineralization. The $ZnCl_2$-like biomineral forms nanoscaled microfibrils and their axial direction points at the top of tooth cusp. The microfibrils aggregate random into intermediate filaments, forming a hierarchical structure. A sketch map of the cricket tooth cusp was proposed and a detailed discussion was given in this paper.

## 1. Introduction

Biomineralization is a quite universal phenomenon in nature, which is frequently found in the claws, shell and teeth of invertebrates and in the bones of vertebrates [1-4]. Calculus is also a typical biomineralization production in mammals. In addition, bamboo and timber were also reported to have biomineralized phenomena [5-6]. Among the rich and colorful biomineralizations, the calcium salts are the most common biominerals. For example, $CaCO_3$ was found in corals and pearls, as well as in the horns, claws, stings, and shells of invertebrates or mollusk. CaP was found in teeth and bones of vertebrates. $CaO_x$ was reported in plants and kidney stones [7-8]. Besides, the biominerals containing Si or Fe are also quite hackneyed. In 2002, a rare copper-based atacamite $[Cu_2(OH)_3Cl]$ was found in the jaws of the marine bloodworm [1]. Evidently, teeth are undoubtedly the most common biominerals no matter in invertebrates or vertebrates.

Cricket, also named as Ququ in Chinese, is artificially feeded as pet because of its arioso stridulation and pugnacious character for a centuries-old history in China. Cricket-fight was a popular distraction in ancient China. The fighting weapons of cricket are a pair of hard and stiff teeth which are also its survival talisman. At present, we found amazedly that a cricket can cut through the 20-μm Al foil in thickness and left a hole with diameter of about 10 mm on the box-wall when it escaped from an Al-foil box. This example illustrates that the sharp teeth of cricket are indeed hard enough. All these stimulate us to clarify whether the cricket teeth is biomineralized or not, as well as what is the possible structure in cricket teeth. In this paper, synchrotron radiation micro-beam X-ray fluorescence (XRF) is used to probe the element concentration and distribution in the cricket teeth. Wide angle X-ray scattering (WAXS) is used to detect the crystalline structure of the possible biominerals in the cricket teeth. Small angle X-ray scattering (SAXS) is used to analyze the scatterer sizes in the cricket teeth. At the same time, scanning electron microscope (SEM) is used to directly observe the nanoscaled structures in the interior of the cricket teeth.

## 2. Experimental procedure

### 2.1 Material

Several crickets were captured out of door. Their teeth were cut down for WAXS, SAXS, XRF and SEM measurements. The photograph of two crickets is shown in Fig. 1(A). A single tooth taken out from a cricket is shown in Fig. 1(B). Roughly, it can be seen that the cricket teeth are just like a pair of scissors, but with serrate edges in the inner sides. The color of the tooth edges, especially, the tooth cusp is deeper, which implies probably the change of the element distribution or concentration.

Received 15 February 2012, Revised 15 February 2012
1) Corresponding author. E-mail: wuzh@ihep.ac.cn
*: Supported by the National Natural Scientific Foundation (10385008), Knowledge Innovation Program of Chinese Academy of Sciences (KJCX3-SYW-N8) and Momentous Equipment Program of Chinese Academy of Sciences (YZ200829)

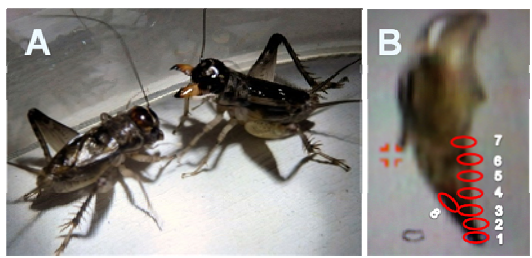

**Fig. 1** Photograph of cricket (A) and the eight locations exposed to incident X-ray are, respectively, marked as 1-8 on the cricket tooth (B).

### 2.2 WAXS and SAXS measurements

WAXS and SAXS experiments were performed at beamline 1W2A of Beijing Synchrotron Radiation Facility (BSRF). The incident X-ray was monochromized to a wavelength of 0.154 nm by a Si (111) triangle bending crystal. A two-dimensional CCD detector was utilized to record the WAXS and SAXS patterns. The cricket tooth was fixed on the sample holder. For the WAXS experiment, the sample-to-detector distance was fixed at 145 mm, covering a scattering angle (2θ) range of 5 - 35°. Total eight locations at the cricket tooth were chosen to detect the WAXS patterns. In which, seven locations were marked as 1-7 from the tooth cusp to the tooth root, and another location was marked as 8 as shown in the Fig. 1(B). For the SAXS experiment, the sample-to-detector distance was fixed at 1265 mm. The tooth locations exposed to incident X-ray and the corresponding SAXS patterns are shown in Fig. 2.

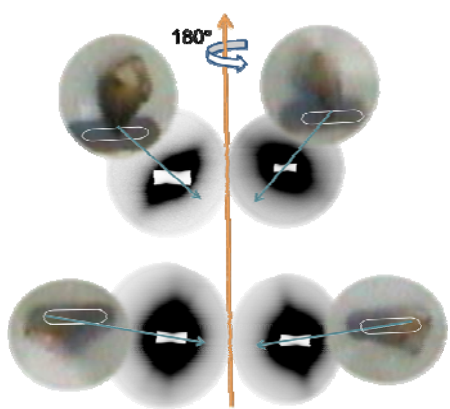

**Fig. 2** Attitude of cricket tooth, location exposed to incident X-ray, and the corresponding SAXS patterns.

### 2.3 XRF measurements

Micro-beam XRF experiments were performed at beamline 4W1B (XRF station) of BSRF. The incident X-ray was focused on the sample with beam size of 50×50 μm$^2$. A Si(Li) detector was used to record the fluorescence radiations came from the sample. This detector was set to be perpendicular to the incident X-ray beam. The normal direction of sample was located between the detector and the incident X-ray, respectively, with an angle of 45°. The incident X-ray energy was fixed at 15 keV by a Si(111) double-crystal monochromator. The XRF experiment was carried on at ambient condition. By scanning the sample step by step, the element distribution and their relative concentration were obtained. For comparison between XRF and WAXS results, the average element concentrations were, respectively, extracted for the eight locations shown in the Fig. 1(B).

### 2.4 SEM observation

In order to observe the nanoscaled structures in the interior of the cricket tooth, the cricket tooth was cut off along its longitudinal direction. A layer of gold nanoparticles with diameter of 3-5 nm was sprayed on the cut section of the tooth. The tooth and sample holder were linked by conductive tape. SEM observation was performed with the sample in vacuum. The SEM micrograph of the cut section is shown in Fig. 3(A). Two selected areas, respectively, from the tooth cusp and tooth body were further amplified for more details as shown in Fig. 3(B) and 3(C). At the same time, the element analysis at the tooth cusp was also given by the energy dispersive X-ray spectroscopy (EDX) as shown in Fig. 3(D).

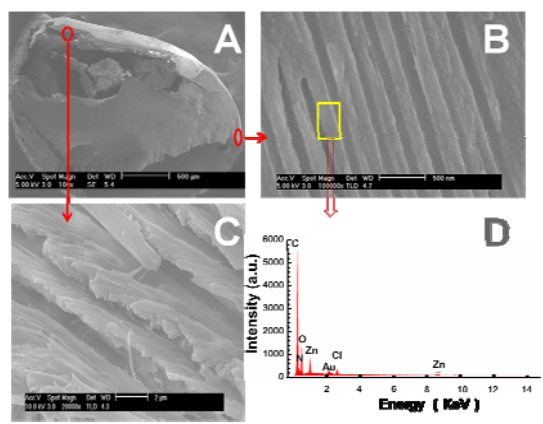

**Fig. 3** SEM micrograph of the cross section of the cricket tooth (A), Selected-area enlarged maps of tooth cusp (B) and tooth root (C), as well as SEM energy dispersive spectrocsopy (D).

## 3. Results and discussion

In order to clarify the possible biomineralization and the corresponding mineral structure in the teeth of cricket, the element distribution and concentration were firstly checked. Synchrotron radiation micro-beam XRF technique and SEM-EDX were used to detect the element fluorescence signals in the cricket tooth. The XRF experimental data were analyzed by using program Winqsas [9]. The relative element contents are shown in Fig. 4(A). It can be seen that Zn and Br are two prominent heavy elements with the highest concentrations. Fe and As are also detectable in the cricket tooth. By comparing the changes of element concentration with the locations in the cricket tooth, we find that the Zn concentration is obviously decreased from location 1 to location 7. This demonstrates that the tooth cusp has the highest Zn concentration, which is tens of times higher at the tooth cusp than at other parts. From the photograph shown in Fig. 1, we can find that the color of the cricket tooth become deeper around the tooth cusp. This result tells us that the color depth is positively correlated with the Zn element concentration. However, the Br concentration is almost constant in all the eight detected locations. Evidently, the constant Br concentration illustrates that Br is a common element in an insect organization. At the meantime, it also implies that Br is not the necessary element in the possible biomineralization of the cricket tooth. We believe that the possible minerals in cricket tooth should be Zn-based compound if the cricket tooth is biomineralized. The higher concentration of Zn element was also confirmed by the EDX experiments for the tooth cusp. Besides the light elements (C, N and O), Zn and Cl are the main heavy elements in the cricket teeth. However, we do not find the contributions of Fe, As, and Br from the EDX. And we do not find the Cl element from XRF measurements. These can be attributed to the different sensibility between XRF and EDX techniques, as well as the different sample status. In XRF measurement, the sample was kept undestroyed. The incident X-ray beam was directly radiated to the sample surface. Because the dimension (~3 mm) of the cricket tooth is larger and the energy (~2.62 keV) of Cl-K$\alpha$ fluorescence is lower, the fluorescence signal of Cl element can be fully absorbed by the tooth and air as Cl locates in the interior of the tooth. In EDX measurement, the tooth was cut off. The incident electron beam was directly radiated to the cut plane of the tooth. And the vacuum environment did not attenuate the intensity of Cl-K$\alpha$ fluorescence. The appearance of Cl-K$\alpha$ fluorescence illustrates that Cl is distributed in the interior of the tooth. The disappearance of Fe, As, Br fluorescence signals in EDX measurements demonstrates that Fe, As, Br are distributed in the surface of the tooth. Moreover, metallic element Zn always exists in the interior and surface of the tooth, which can increase the hardness of teeth [10].

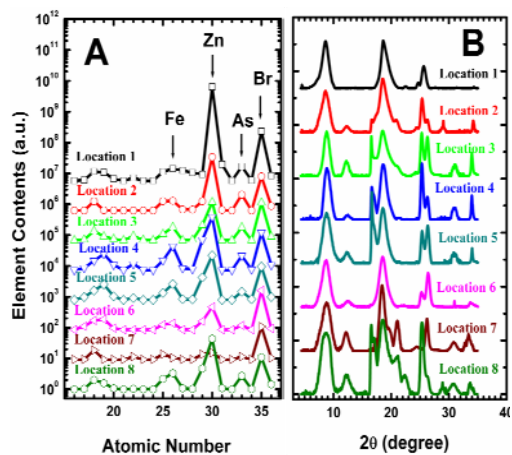

**Fig. 4** Element contents of the eight locations exposed to incident X-ray (A) and the corresponding WAXS patterns (B).

If the cricket tooth was biomineralized, then what is the possible biominerals in the cricket teeth? In order to get the possible crystalline structures, WAXS technique was used to collect the diffraction patterns, respectively, at the eight locations of the tooth. These 2D WAXS diffraction patterns were transferred into 1D WAXS curves by using Fit2d program [11]. After removal of the background, the 1D WAXS curves are shown in Fig. 4(B) for all the eight locations in the cricket tooth. It can be seen that the eight locations all have obvious diffraction peaks. Roughly, the WAXS data are quite similar, but a difference between location 1 and the others is also easy to be found. That is to say, the diffraction peaks of location 1 seem to come from a single phase, while the diffraction peaks of the other locations are possibly the contributions of multiphase. Evidently, these diffraction peaks of location 1 appear always on the other locations, here we attribute temporarily them to phase I. Besides the diffraction peaks from phase I, the extra diffraction peaks appeared on the other locations are ascribed to phase II. Comparatively speaking, the diffraction peaks of phase I are wider than phase II, which means that the grain sizes are smaller or the structural disorder is larger in phase I than in phase II. Generally, all diffraction peaks are quite broadened because of the structural incompleteness of crystals or the smaller grain sizes in the cricket tooth. At the same time, the angular extent (2$\theta$) with visible diffraction peaks is only from 5° to 35°. All these limit us to do a precise Rietveld refinement for the diffraction data. In order to give the possible crystal structures in the cricket tooth, the program FindIt [12] were used to search for the known structures containing elements Zn, Fe, As and Cl from the Inorganic Crystal Structure Database (ICSD). By comparing the known structures with the experimental diffraction data, it is found the diffraction data can not fitted well by only one compound. In fact, Br and Cl are, respectively, distributed in the surface and interior of the tooth as discussed above. Two compounds $ZnFe_2(AsO_4)_2(OH)_2(H_2O)_4$ [13] (Ojuelaite mineral) and $ZnCl_2$ [14] (code 15918 in ICSD) together have a best coincidence between the experimental data and the known structures as shown in Fig. 5. According to the above discussion, we think that a compound like $ZnFe_2(AsO_4)_2(OH)_2(H_2O)_4$ is from the surface of the tooth and is named as surface phase, while a compound like $ZnCl_2$ is from the interior of the tooth and is named as interior phase. Here, we have to say that the diffractions of location 1 are not only from the surface phase. Because location 1 is corresponding to the tooth cusp, the surface content is absolutely higher than the interior content. Therefore, interior phase is relatively weak and is not easy to be distinguished from the diffractions. Although the crystalline structures in the cricket tooth were not obtained by a whole pattern fitting, we still predict two possible crystal structures by comparing the diffraction patterns. Perhaps, the surface phase is common characteristic, but the interior phase can be ascribed to the biomineralization of the cricket teeth.

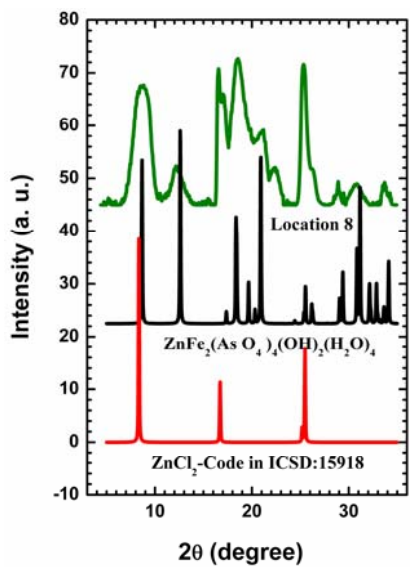

**Fig. 5** Comparison between the experimental WAXS pattern of location 8 and the diffraction data of two optimum crystalline phases.

To get more details about the cricket tooth, SEM was used to observe the inner configuration of the cricket tooth as shown in Fig. 3(A). It can be seen that the cricket tooth has a relatively stout shell and a relatively soft interior filled with stroma. Two selected areas, respectively, at the tooth cusp and the tooth root were further amplified as shown in Fig. 3(B) and 3(C). It can be seen that the tooth root consists of stratified structures, while the tooth cusp contains a lot of parallel fibrils with diameter of about 150-200 nm. The elements in the tooth cusp was also detected by EDX as shown in Fig. 3(D), the detected area is shown in Fig. 3(B) as a rectangle. We believe that it is the stratified structures make the shell of cricket tooth hard, and the parallel fibril structures make the tooth cusp stiff.

In order to clarify the nanoscaled structure in the cricket tooth, SAXS measurements were performed for the tooth cusp and the tooth body twice, respectively. The difference in the two measurements for the tooth cusp and tooth body is that the sample was turned 180° with the incident X-ray radiating to the tooth from the front or from back directions. The 2D SAXS patterns and the sample's positions were shown in Fig. 2. The obvious anisotropy of the SAXS patterns illustrates that there are preferred orientation of scatterers in the tooth. The short axis in the SAXS pattern is along the fibrils' growing direction, and the long axis in SAXS pattern is along the fibrils' cross section. When the sample was turned 180°, roughly the SAXS pattern was also turn 180°. Therefore, we can judge that the long axis of the fibres in the tooth points at the direction of tooth cusp.

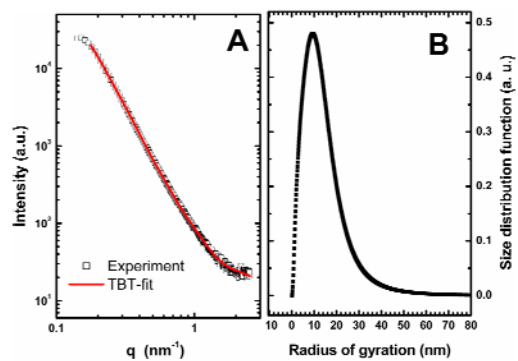

**Fig. 6** Experimental SAXS intensity (open square) and its fitting curve (solid line) with TBT method (A), as well as size distribution function of the scatterers in the cricket tooth (B).

The scattering signal along the long axis of SAXS pattern was extracted with Fit2D program as shown in Fig. 6(A) and marked as open squares. The TBT method [15] was used to analyze the size distribution of scatterers. The calculated SAXS curve (marked as solid line) is in excellent agreement with the experimental one as also shown in Fig. 6(A). Based on the SAXS data along the long axis of SAXS pattern, the radius of gyration distribution of fibrils in tooth is obtained as shown in Fig. 6(B). It can be seen that the most probable value of the radius of gyration is around 9.4 nm, but the average value is about 12.1 nm. Based on a wafer model, the most probable diameter and the average one of the fibers are, respectively, 26.6 and 34.2 nm. Evidently, these values are still smaller than the SEM observed values (~150 nm) as shown in Fig. 3(B). Therefore, we define the fibers shown on SEM micrograph as intermediate filament, and the fibers measured by SAXS as microfibril. That is to say, the intermediate filament consists of lots of microfibrils. Because we did not observe the diffraction in SAXS patterns, the microfibrils should aggregate random into intermediate filament.

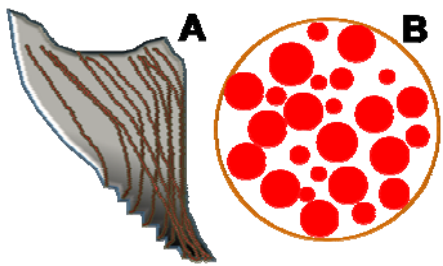

**Fig. 7** Sketch maps of the intermediate filaments in the cricket tooth (A) and microfibrils aggragation into intermediate filament (B). The average diameters of the intermediate filaments and the microfibrils are, respectively, about 150 and 34 nm.

On the basis of SAXS and SEM results, a sketch map of the interfilament in the cricket tooth cusp is proposed and shown in Fig. 7(A). The microfibrils aggregation into intermediate filament is schematically shown in Fig. 7(B). In fact, the fibrous structures [16-18] of biominerals were also found in the enamel of vertebrates' teeth, sea urchin, mollusk and marine bloodworm teeth. These mineral fibrils are beneficial to raising the chewing ability of teeth. They are also helpful for invertebrates to scrape algae off of rocks [19]. Indeed, nanoscaled mineral fibrils strengthen the wear-resisting capability of teeth [20]. Similarly, the hierarchical structures consisted of nanoscaled microfibrils and the biomineralization strengthen the cricket teeth so that the teeth become a truculent weapon besides chewing function.

## 4. Conclusions

The structures of cricket teeth have been studied by using synchrotron radiation XRF, XRD, and SAXS techniques, as well as SEM observation. It has been found that zinc is the main metallic element. The tooth was found to have a surface phase and an interior phase. By XRD data analysis, $ZnFe_2(AsO_4)_2(OH)_2(H_2O)_4$ and $ZnCl_2$ were found to be the most appropriate candidates, respectively, of the surface and interior phases in cricket teeth. The tooth cusp was found to consist of microfibrils with their axial direction pointing at the top of tooth cusp. In addition, the microfibrils aggregate random into intermediate filaments, forming a hierarchical structure with nanoscaled dimension in the tooth cusp. It was the appearance of Zn elements as well as the fibrous and hierarchical structures that strengthen the cricket teeth, which make the cricket as truculent insect with its hard and sharp teeth as weapon.

---